\documentclass[aps,prl,floatfix,twocolumn]{revtex4}
\usepackage{graphicx}
\usepackage{dcolumn}
\usepackage{dcolumn}
\usepackage{amssymb}
\usepackage{amsmath}

\begin{document}

\title{Effective-Hamiltonian modeling of external pressures in
  ferroelectric perovskites}

\author{Jorge \'I\~niguez, J. B. Neaton, and D. Vanderbilt}

\affiliation{Department of Physics and Astronomy,Rutgers University,
Piscataway, New Jersey 08854-8019, USA}

\begin{abstract}
The phase-transition sequence of a ferroelectric perovskite such as
BaTiO$_3$ can be simulated by computing the statistical mechanics of a
first-principles derived effective Hamiltonian [Zhong, Vanderbilt and
Rabe, Phys. Rev. Lett. 73, 1861 (1994)]. Within this method, the
effect of an external pressure (in general, of any external field) can
be studied by considering the appropriate ``enthalpy'' instead of the
effective Hamiltonian itself. The legitimacy of this approach relies
on two critical assumptions that, to the best of our knowledge, have
not been adequately discussed in the literature to date: (i) that the
zero-pressure relevant degrees of freedom are still the only relevant
degrees of freedom at finite pressures, and (ii) that the truncation
of the Taylor expansion of the energy considered in the effective
Hamiltonian remains a good approximation at finite pressures. Here we
address these issues in detail and present illustrative
first-principles results for BaTiO$_3$. We also discuss how to
construct effective Hamiltonians in cases in which these assumptions
do not hold.
\end{abstract}

\maketitle

\section{I. Introduction}

A few years ago it was shown that it is possible to reproduce the
phase transition sequence of a ferroelectric such as BaTiO$_3$ by
simulating the statistical mechanical properties of an effective
Hamiltonian built on the basis of first-principles
calculations~\cite{zho94}. Since then, this approach has been used to
study the phase diagrams and electromechanical responses of other
perovskite crystals (SrTiO$_3$~\cite{zho95b}, PbTiO$_3$~\cite{wag97},
and KNbO$_3$~\cite{kra99}), and even of {\it disordered} perovskite
solid solutions (Pb(Zr$_{1-x}$Ti$_x$)O$_3$ (PZT)~\cite{bel00} and
Pb(Sc$_{0.5}$Nb$_{0.5}$)O$_3$~\cite{hem00}). These calculations have
typically resulted in a very good qualitative agreement with
experiment and clear physical pictures of the phenomena
studied. Presently, even though the method usually gives poor results
for the actual values of the transition temperatures and other
quantities, it is regarded as having the predictive power necessary to
tackle materials-design problems~\cite{ini01,geo01}.

An effective Hamiltonian can be viewed as a Taylor expansion of the
energy of the system, in terms of a set of relevant degrees of
freedom, around a high-symmetry reference structure. For ferroelectric
perovskites, the relevant variables are typically the strains and,
most importantly, the local polar distortions that sum up to produce
the spontaneous polarization. The high-symmetry reference structure is
chosen to be the non-polar, cubic perovskite phase, whose equilibrium
volume is computed {\it ab initio}.

Very conveniently, within this approach it is straightforward to
consider the effect of external fields on the system. One just has to
add a term to the Hamiltonian that transforms the energy into the
appropriate enthalpy. For instance, an external hydrostatic pressure
$p$ would enter the model in a term $pV$, where $V$ is the volume of
the simulated system. Examples of such applications are the
calculation of the $T$-$p$ phase diagram of BaTiO$_3$~\cite{zho95a}
and, very recently, the study of electric-field driven transition
paths in PZT~\cite{bel01}.

This approach to modeling external fields stems directly from
statistical mechanics, and it is exact {\it provided the considered
Hamiltonian is the complete Hamiltonian of the system}. Using
effective {\it incomplete} Hamiltonians actually involves a number of
implicit approximations in the description of the effect of an
external field. To the best of our knowledge, these implicit
approximations have neither been adequately discussed, nor has their
importance been carefully tested, in the literature to date.

In this paper we address these questions in detail. For concreteness,
we consider the case of an external pressure, but our formal argument
applies to any external field. We illustrate the discussion with
calculations of the effect of pressure on BaTiO$_3$, which should be
representative of ferroelectric perovskites and is specially
interesting because of the discrepancy between the
theoretical~\cite{zho95a} and experimental~\cite{ish97} results for
its $T$-$p$ phase diagram.

The paper is organized as follows. In Section~II we discuss the
pressure dependence of the relevant degrees of freedom that define the
effective Hamiltonian. In Section~III we show how the truncation in
the effective-Hamiltonian energy expansion actually implies an
approximation in what we call the ``pressure dependence'' of the
parameters of the model. In Section~IV we tackle the issue of the
anharmonic couplings between the chosen relevant degrees of freedom
and the rest of variables in the system. Finally, we summarize and
present our conclusions in Section~V.

\section{II. Relevant degrees of freedom as a function of pressure}

The fundamental variables entering effective Hamiltonians for
ferroelectric perovskites are the localized, polar displacement
patterns associated with the spontaneous polarization. In order to
define these local polar modes {\it ab initio}, one examines the
calculated phonon~\cite{fn1} dispersion curves of the high-symmetry
cubic phase; the soft phonons (i.e., those with very small or negative
force constant) are obviously those to be included in the model. (Note
that the spontaneous polarization will correspond to an unstable
zone-center phonon.) The local modes, or lattice Wannier functions,
are then obtained from the corresponding eigenvectors~\cite{fn:wan}.

Taking the calculated equilibrium, high-symmetry phase as a reference
implies that the relevant variables we determine are actually the
relevant variables {\it at zero pressure}. Including the effect of
pressure by simply adding an extra $pV$ term to the Hamiltonian thus
relies on the fundamental assumption that the relevant variables at
zero pressure will remain the relevant variables at finite
pressure. There are two situations in which this assumption could
fail, namely the applied pressure could give raise to additional soft
modes and/or the zero-pressure relevant configuration space could
change significantly. In the former case, the obvious solution is to
include the new soft modes in the model. In the latter, one would have
to redetermine the relevant polar degrees of freedom, now under
applied pressure, and recalculate the effective Hamiltonian
accordingly.

We have performed first-principles calculations to study these issues
in BaTiO$_3$~\cite{fn2}. In this material, the modes that most likely
may become soft under compression are those at the M and R
zone-boundary points that involve rotations of the oxygen
octahedra. We have calculated the evolution of the eigenvalues of such
modes as a function of the lattice parameter, ranging from
$a=7.46$\,a.u.\ (the calculated equilibrium lattice parameter of the
cubic phase) to $a=7.36$\,a.u.\ (which is well beyond the value at
which the ferroelectric instability disappears; see next section). We
find no softening with increasing pressure; on the contrary, the modes
become slightly harder. (For instance, the eigenvalues of the modes at
R, which are softer than those at M, are 0.0200\,a.u.\ and
0.0220\,a.u.\ for $a=7.46$\,a.u.\ and $a=7.36$\,a.u.\ respectively.)
On the other hand, the evolution with pressure of the relevant
variables related to the ferroelectric instability can be monitored by
observing the change of the zone-center soft mode. We find that the
overlap between the unstable eigenmode at $a=7.46$\,a.u. and the
corresponding, nearly-unstable eigenmode at $a=7.36$\,a.u. is above
$98\%$.  In summary, for BaTiO$_3$ the relevant variables defined at
the zero-pressure equilibrium lattice constant remain a very good
approximation all through the interesting pressure range.

\section{III. ``Pressure dependence'' of the Hamiltonian parameters}

Let us assume that the zero-pressure relevant degrees of freedom
continue to be the relevant ones in the pressure range of interest. In
the following we argue that, even in this case, the effect of pressure
is not fully captured by the addition of the $pV$ term.

For simplicity, we consider a one dimensional ferroelectric, and
describe it in terms of the macroscopic variables $P$ (polarization)
and $\eta$ (strain). For small $P$ and $\eta$, the energy of the
system can be written as a low-order Taylor expansion around the
high-symmetry paraelectric phase ($P=\eta=0$),
\begin{equation}
E(P,\eta) = E_0 - C_{2,0} P^2 + C_{4,0} P^4 + C_{0,2} \eta^2 - C_{2,1}
P^2 \eta,
\label{eq:energy}
\end{equation}
where all the expansion coefficients (in obvious notation) are chosen
to be positive. (The maximum orders included in the expansion of
Eq.~(\ref{eq:energy}) are those normally adopted in the effective
Hamiltonians proposed to date in the
literature~\cite{zho94,kra99,bel00,hem00}.) Let us include the effect
of an external pressure $p$ by adding the term $pV_0(1+\eta)$, with
$V_0$ the equilibrium volume of the paraelectric phase. It is now
convenient to consider the change of variables
\begin{equation}
\eta \rightarrow \widetilde{\eta} - \frac{V_0}{2C_{0,2}}p \equiv
\widetilde{\eta} + \eta_p
\label{eq:varchange}
\end{equation}
where $V_0(1+\eta_p)$ is the equilibrium volume of the paraelectric
phase under applied pressure. The enthalpy then becomes
\begin{eqnarray}
E(P,\eta;p) & = & E_0(p) - \left[ C_{2,0} - \frac{V_0 C_{2,1}}{2
C_{0,2}}p\right] P^2 \nonumber \\
& & + C_{4,0} P^4 + C_{0,2} \widetilde{\eta}^2 - C_{2,1} P^2
\widetilde{\eta},
\label{eq:enthalpy0}
\end{eqnarray}
where $E_0(p)$ is the pressure-dependent energy of the non-polar
phase. If we now denote the bracketed prefactor of the $P^2$ term as
$\widetilde{C}_{2,0}$, and also introduce $\widetilde{E}_0$, we can
rewrite the enthalpy as
\begin{equation}
\widetilde{E}(P,\widetilde{\eta}) = \widetilde{E}_0 -
\widetilde{C}_{2,0} P^2 + C_{4,0} P^4 + C_{0,2} \widetilde{\eta}^2 -
C_{2,1} P^2 \widetilde{\eta},
\label{eq:enthalpy}
\end{equation}
which is formally identical to Eq.~(\ref{eq:energy}). Thus, we have
proven that, within our model, the system under applied pressure is in
all respects equivalent to the original, pressure-free system
described by Eq.~(\ref{eq:energy}), but with some modified parameters
(indicated by tildes) that can be regarded as being ``pressure
dependent''. (In the following, references to the ``pressure
dependence'' of the parameters of the Hamiltonian should be understood
in this sense.)

Having reformulated the problem in this way, an obvious question
arises.  If $E_0$ and $C_{2,0}$ are modified by pressure, why is it
that the remaining parameters in Eq.~(\ref{eq:energy}) are not? The
answer is that pressure should indeed affect the rest of the
parameters in Eq.~(\ref{eq:energy}), but this effect is not captured
by our model because we are using a truncated Taylor expansion of the
energy. Imagine, for instance, that Eq.~(\ref{eq:energy}) contained
the higher-order term $C_{4,1} P^4 \eta$. The variable change in
Eq.~(\ref{eq:varchange}) would then generate new terms $C_{4,1} P^4
\widetilde{\eta} + C_{4,1} \eta_p P^4$, thus modifying $C_{4,0}$ to
become $\widetilde{C}_{4,0} = C_{4,0} + C_{4,1}
\eta_p$.

The previous argument can readily be applied to the microscopic
effective Hamiltonians of interest~\cite{fn3}. We have thus shown that
a truncation of the energy expansion, which may be perfectly well
justified for the calculation of zero-pressure properties, actually
implies an approximation in the pressure dependence of the parameters
of the model. Obviously, we can improve on this approximation by
including higher-order terms in the effective Hamiltonian. However,
our experience tells us that it is usually difficult to calculate
high-order terms accurately, and the very issue of where to truncate
the expansion may become a problem. There is, fortunately, an
alternative approach suggested by the derivation above. One can
instead take the equilibrium, high-symmetry phase at pressure $p$ as
the reference structure from which one calculates the Taylor expansion
of the energy. Such an expansion, which could likely be kept
low-order, would be analogous to Eq.~(\ref{eq:enthalpy}) but with all
the parameters containing their full pressure dependence by
construction.

\begin{figure}[t!]
\includegraphics[angle=-90,width=\columnwidth]{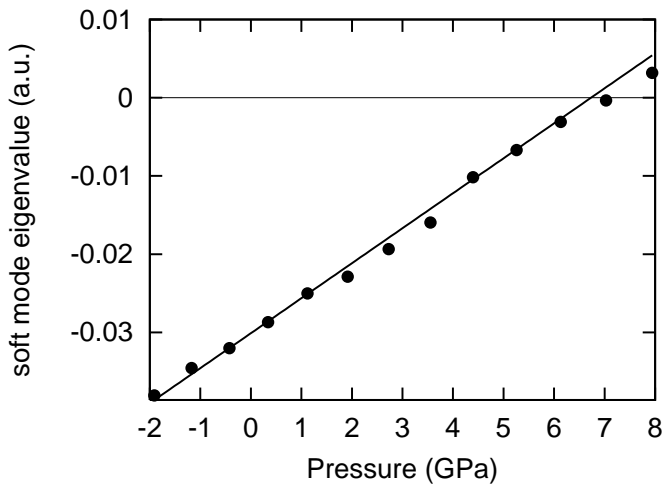}
\caption{Calculated eigenvalue of the 
ferroelectric soft mode of BaTiO$_3$ as a function of pressure (dots),
as compared with the behavior predicted (see text) from the
zero-pressure effective Hamiltonian (solid
line). (See~\protect\cite{fn2} for the technical details of the {\it
ab initio} calculations.) From the explicit calculation we obtain a
ferroelectric transition point at $p_c = 7.0$\,GPa ($V_c/V_0 =
0.966$), while the effective-Hamiltonian approach places the
transition at $p_c = 6.7$\,GPa ($V_c/V_0 = 0.964$, as given by the
equation of state built into the effective Hamiltonian itself). In
both cases, the calculated equilibrium volume of the cubic phase is
$V_0 = 414$\,a.u.}
\label{fig:C20}
\end{figure}

In order to make these ideas explicit, we have calculated the pressure
dependence of the parameters in the effective Hamiltonian of
BaTiO$_3$~\cite{fn2}. The soft-mode eigenvalue, which is exactly the
equivalent of $-2\widetilde{C}_{2,0}$ in Eq.~(\ref{eq:enthalpy}), is
computed as a function of the unit-cell volume, and the corresponding
pressure is determined from the equation of state of the cubic
phase. The results are shown in Fig.~\ref{fig:C20}. We find that the
linear law resulting from the $pV$ term (see the bracketed term in
Eq.~(\ref{eq:enthalpy0})) turns out to agree surprisingly well with
the values explicitly calculated at different pressures. We also
computed the pressure dependence of the parameters that are analogous
to $C_{4,0}$, $C_{0,2}$ and $C_{2,1}$ in Eq.~(\ref{eq:energy}). In all
cases the values calculated as a function of pressure lay within
approximately $10\%$ of the zero-pressure value. Very importantly,
this applies to the parameter that determines the relative stability
of the different ferroelectric phases of BaTiO$_3$ ($\gamma'$ in
Ref.~\cite{kin94}).  Therefore, we find that the ground state of the
system is rhombohedral throughout the pressure range in which the
ferroelectric instability exists, in agreement with the
first-principles effective-Hamiltonian results of Zhong {\it et
al.}~\cite{zho95a}.

\section{IV. The coupling between relevant and irrelevant variables}

The experimental work of Ref.~\cite{ish97} suggests that, along
isotherms at very low temperature, BaTiO$_3$ undergoes a transition
sequence with increasing pressure that progresses from the
zero-pressure ferroelectric rhombohedral phase to ferroelectric
orthorhombic, then ferroelectric tetragonal, and finally paraelectric
cubic phases. As we have seen in the previous sections, we have found
no indication of such a behavior in our first-principles results. This
disagreement is not necessarily a serious one, since the authors of
Ref.~\cite{ish97} attributed their observed transition sequence to
quantum fluctuations (zero-point atomic motion), an effect that is not
taken into account in our calculations.  In any case, it led us to
explore yet another possible effect of pressure that could, in
principle, account for such a discrepancy.

The effective-Hamiltonian approach relies on the assumption that the
coupling between the chosen relevant variables and the rest of the
degrees of freedom in the system is not important. Since the relevant
modes are defined in terms of phonons that diagonalize the harmonic
part of the total Hamiltonian of the system, such a coupling is
anharmonic, which partly justifies our neglecting it. However, we know
that in BaTiO$_3$, as we approach the critical pressure at which the
ferroelectric instability disappears, the energy differences among the
various low-symmetry phases go to zero. Small anharmonic effects could
then be essential to the determination of the ground state of the
system. Note that, if this were the case, the usual
effective-Hamiltonian approach, in which the relevant variables are
identified by looking at the {\it harmonic} phonon dispersion curves,
should be modified.

\begin{figure}[t!]
\includegraphics[width=\columnwidth]{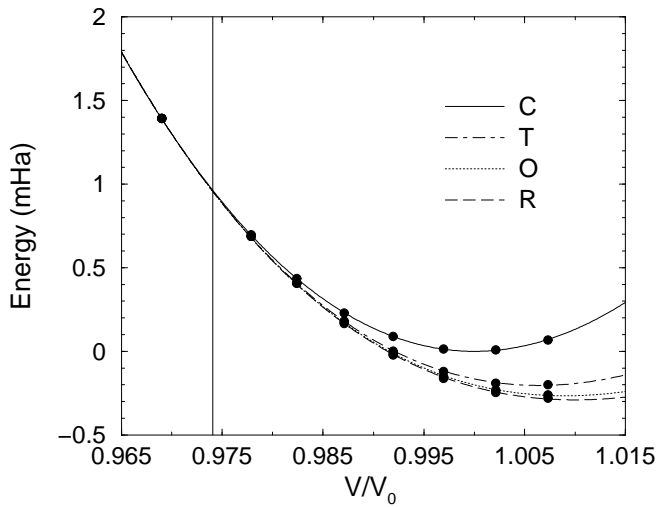}
\caption{Energies of the different phases of BaTiO$_3$ (cubic,
tetragonal, orthorhombic, and rhombohedral) as a function of the
volume of the unit cell. The energies are computed by relaxing the
atomic positions and cell shape under the constraints of preserving
the corresponding symmetry and keeping the cell volume
fixed. (See~\protect\cite{fn4} for the technical details.) We have
marked with a vertical line the ferroelectric transition point between
the non-polar cubic and polar rhombohedral phases, located at $V_c/V_0
= 0.974$ ($p_c = 5.6$\,GPa) where $V_0 = 417$\,a.u.\ is the calculated
equilibrium volume of the cubic phase. The discrepancy between these
numbers and those in the caption of Fig.~\protect\ref{fig:C20} is not
surprising if we take into account that they have been computed using
two different codes (and two different kinds of pseudopotential
schemes).}
\label{fig:EvsV}
\end{figure}

We can study this issue very easily in BaTiO$_3$ by fully relaxing the
structure of the system in different constrained symmetries, while
keeping the volume fixed~\cite{fn4}. The symmetries we consider are of
course the tetragonal, orthorhombic, and rhombohedral ones
corresponding to the ferroelectric phases of the material, and we
explore the interesting range of volumes.  Figure~\ref{fig:EvsV} shows
the energy of the relaxed structures as a function of volume. For
$V/V_0 < 0.980$ the energies of the different phases differ by less
than 0.01\,mHa. Such small differences are beyond the accuracy of our
computational technique and, therefore, we are not able to study the
critical region in detail. Nevertheless, we can determine the critical
volume quite reliably as the point at which $C_{2,0}$ passes through
zero; we obtain $V_c/V_0=0.974$, indicated with a vertical line in
Fig.~\ref{fig:EvsV}. On the other hand, we find that, down to
$V/V_0\sim 0.977$, there is no common tangent among the curves in
Fig.~\ref{fig:EvsV} that would correspond to a first-order phase
transition between two ferroelectric phases at $p < p_c$. It thus
seems reasonable to conclude that the calculated ground state is
rhombohedral up to $p_c$, the point at which it undergoes a
second-order transition to the paraelectric cubic phase. This result
is in complete agreement with the predictions of the
effective-Hamiltonian approach.

The importance of the above-mentioned anharmonic couplings in
determining the ground state of the system can be quantified, for a
given volume, by expanding the displacement vector corresponding to
the relaxed atomic positions as a linear combination of the
zone-center modes calculated at that same volume. We find that,
throughout the interesting range, the ferroelectric soft mode accounts
for approximately $99\%$ of the total atomic relaxation. This result
further justifies neglecting the anharmonic couplings among relevant
and irrelevant variables in the case of BaTiO$_3$.

\section{V. Summary and Conclusions}

In the usual approach to simulating external pressures within the
effective-Hamiltonian scheme, one starts from the calculated
zero-pressure effective Hamiltonian and adds to it a $pV$ term to
obtain the appropriate enthalpy. Such a procedure implies a number of
approximations that we have discussed in detail in this paper. Our
discussion is actually general, and applies to any external field.

On the one hand, we have seen that there are approximations related
with the determination of the relevant degrees of freedom to be
included in the model.  In principle, one should determine the
relevant variables as a function of pressure. One should also check
that the relevant variables do not have significant anharmonic
couplings with the rest of the degrees of freedom of the system
throughout the interesting pressure range.

On the other hand, we have discussed a rather subtle implicit
approximation that has to do with the truncation of the energy
expansion in the effective Hamiltonian.  We have shown that such a
truncation, which may be justified for the purpose of zero-pressure
calculations, actually implies an approximation in what we have called
the ``pressure dependence'' of the parameters of the model. As a
solution to this problem we propose the calculation of effective
Hamiltonians ``under pressure,'' meaning that for a given pressure we
would take as our reference structure the calculated, equilibrium
high-symmetry phase at that pressure.

We have studied in detail all these issues for BaTiO$_3$. We have
shown that all the previous approximations are actually highly
accurate for this material, which we believe should be representative
of ferroelectric perovskites. The usual approach to simulating
external pressures thus seems to be fairly reliable.

Regarding the $T$-$p$ phase diagram of BaTiO$_3$, our results indicate
that at a {\it classical} level (i.e., without considering the
zero-point motion of the ions) the only possible polar phase at 0~K is
the rhombohedral one. On the other hand, as Fig.~\ref{fig:EvsV} shows,
the differences in energy among the ferroelectric phases become very
small as the volume decreases towards $V_c$.  It thus seems possible
that, at low temperatures, the quantum fluctuations of the ions may be
able to change the relative stability of the ferroelectric phases and
lead to a phase diagram similar to the one reported by Ishidate {\it
et al.} (see Fig.~1 in Ref.~\cite{ish97}). This very interesting
question will be addressed elsewhere~\cite{ini_to_come}.

\vskip 2mm

We thank Alberto Garc\'{\i}a and Laurent Bellaiche for their help
regarding various technical issues. This work was supported by the ONR
through Grant N0014-97-1-0048.


\begin{thebibliography}{99}
\bibitem{fn2} For these
calculations we used density-functional theory within the
pseudopotential approach (using Vanderbilt's ultrasoft
pseudopotentials~\cite{van90}) and the local density approximation. We
used a 25~Rydberg cutoff for the plane-wave basis and a $6\times
6\times 6$ Monkhorst-Pack special k-point mesh. Importantly, in our
calculations we considered 10 valence electrons for Ba ($5s^2 5p^6
6s^2$), 12 for Ti ($3s^2 3p^6 3d^2 4s^2$), and 6 for O ($2s^2
2p^4$). These are well-converged calculation conditions, as discussed
in Ref.~\protect\cite{kin94}.
\bibitem{fn3} In the typical effective
Hamiltonian, where the strain is three-dimensional, the $pV$ factor
actually produces a term analogous to $\widetilde{C}_{0,2}$ as well as
other strain-related changes. However, the high-order terms associated
with the polar modes, and the terms that couple them with the strains,
are not changed.
\bibitem{fn4} We performed these
calculations using the Vienna Ab-initio Simulation Package
(VASP)~\protect\cite{vasp}. We used a $6\times 6\times 6$ special
k-point grid and a 36~Rydberg cutoff for the plane-wave basis, and we
worked within the local density approximation. We used the Projector
Augmented Wave (PAW) method~\protect\cite{paw}, and valence electrons
as indicated in~\protect\cite{fn2}.
\end{thebibliography}
\end{document}